\documentclass[prb,reprint]{revtex4-2}
\usepackage{graphicx}
\usepackage{enumitem}
\usepackage{amsmath}
\newcommand{\ket}[1]{|{#1}\rangle}

\usepackage{amsfonts}

\usepackage[usenames,dvipsnames]{xcolor}
\usepackage[colorlinks=true,citecolor=Cerulean,linkcolor=RubineRed,urlcolor=Cerulean]{hyperref}

\begin{document}

\title{Critical states and anomalous mobility edges in two-dimensional diagonal quasicrystals}

\author{Callum W. Duncan}
\email{callum.duncan@strath.ac.uk}
\affiliation{Department of Physics and SUPA, University of Strathclyde, Glasgow G4 0NG, United Kingdom}

\begin{abstract}
We study the single-particle properties of two-dimensional quasicrystals where the underlying geometry of the tight-binding lattice is crystalline but the on-site potential is quasicrystalline. We will focus on the 2D generalised Aubry-Andr\'e model which has a varying form to its quasiperiodic potential, through a deformation parameter and varied irrational periods of cosine terms, which allows a continuous family of on-site quasicrystalline models to be studied. We show that the 2D generalised Aubry-Andr\'e model exhibits single-particle mobility edges between extended and localised states and a localisation transition in a similar manner to the prior studied one-dimensional limit. However, we find that such models in two dimensions are dominated across large parameter regions by critical states. The presence of critical states results in anomalous mobility edges between both extended and critical and localised and critical states in the single-particle spectrum, even when there is no mobility edge between extended and localised states present. Due to this, these models exhibit anomalous diffusion of initially localised states across the majority of parameter regions, including deep in the normally localised regime. The presence of critical states in large parameter regimes and throughout the spectrum will have consequences for the many-body properties of quasicrystals, including the formation of the Bose glass and the potential to host a many-body localised phase. 
\end{abstract}
\pacs{}

\maketitle

\section{Introduction}

Quasicrystalline systems are an intriguing intermediary between periodic crystalline and amorphous disordered systems \cite{Shechtman1984,janssen2018,berger2000}, characterised by the presence of long-range order and short-range disorder. There has been extensive study of their electronic properties by considering one-dimensional quasiperiodic models including interactions \cite{Roux2008,Deng2008} where the quasiperiodic nature of the system is included within the parameters and vertex models of aperiodic tilings \cite{Tsunetsugu1991,Rieth1995,Repetowicz1998,Prunele2002}, which largely probe the physics due to their quasicrystalline geometry. One of the most intriguing properties of quasicrystals is their potential to host a many-body localised phase in two dimensions, due to the conjecture that there is an absence of rare regions \cite{Iyer2013,Gopalakrishnan2016,agarwal2017,Doggen2019,mace2019}. 

The physical properties of many-body quasicrystalline systems is an open question, especially beyond one dimension,  due to the complexity of theoretically simulating two-dimensional many-body systems. The quasicrystalline nature of the system often leads to the presence of frustration due to short-range disorder, while long-range order means that large system sizes are required to probe physical properties. This is potentially an area where quantum simulators, as have been pursued for 2D quasicrystals with cold atoms \cite{Sanchez2005,Viebahn2019Matter,Sbroscia2929Observing,yu2023observing}, could have a significant impact. This has inspired a number of recent works both looking at Hubbard-type lattice models \cite{Johnstone2019,Ghadimi2020,Johnstone2021Mean,johnstone2022barriers,Gottlob2023} and continuous quantum Monte-Carlo \cite{Gautier2021,Ciardi2022,Zhu2023} for ultracold atom in optical quasicrystalline lattices. However, in order to fully understand the physical interplay of quasiperiodicity and interactions, we first need to further develop our understanding of single-particle quasicrystalline models in 2D.

A typical example of a tight-binding quasicrystal is the one-dimensional Aubry-Andr\'e (AA) model, which has a transition between extended and localised states and is described by the sampling of a cosine term of irrational period with the spacing of a tight-binding model \cite{Harper1955Single,aubry1980analyticity,dominguez2019aubry}.  It is known that the one-dimensional AA model does not have a mobility edge due to the localisation transition being defined by a self-dual point. However, it can be deformed to allow for the existence of mobility edges and a mixed intermediary regime between that of extended and localised states in a related family of quasiperiodic potentials \cite{Ganeshan2015Nearest,Liu2015,Li2015Many,monthus2019multifractality,Li2020Mobility,He2022Persistent}, this is called the Generalised AA (GAA) model.  Recently, the 1D GAA model has been explored in experimental realisations utilising ultracold atoms \cite{An2021Interactions,Wang2022} which included the interplay of interactions and quasicrystalline order. A related realisation of the 1D AA model was also recently studied using cavity polaritons \cite{goblot2020emergence}, in this case the experimental potential mapped between the AA and Fibonacci chains to explore a family of quasicrystalline models.

An interesting property of quasicrystals is that they can host critical states away from transition regions.  Critical states are characterised by being neither fully localised or extended in the space that supports them and are multifractal in the Hilbert space. When moving to two-dimensional models the role of critical states becomes important. This has been well-studied in vertex models of quasicrystals \cite{grunbaum1987,cockayne2000,Baake2002,Grimm2003Energy,Mace2017Critical}, which are quasiperiodic through a variation of the local coordination number with the on-site term normally set to zero.  It is known that two-dimensional generalisations of the AA model can host partially extended states, which are critical states, due to weak modulation lines in the potential \cite{Szabo2020Mixed}. It has also recently been observed that some one-dimensional quasicrystals can exhibit anomolous mobility edges between localised and critical states in the spectrum \cite{Liu2022Anomalous}. The main motivation of this work is to extend upon these prior observations and build a picture of the single-particle behaviour of 2D quasicrystalline models,  where the quasiperiodic term is from the on-site component of a tight-binding Hamiltonian, with a particular focus on the role of critical states.  While we will focus on the single-particle picture, some of the impact of the critical states in the many-body regime have already been probed by the consideration of superfluid and transport properties that are supported by the weak modulation of the potential in certain regions \cite{Johnstone2021Mean,johnstone2022barriers,strkalj2022}.

In this work we will first define the 2D GAA model and the measures that will be utilised to distinguish localised, extended, and critical states. We will then show in Sec.~\ref{sec:Properties} that the 2D GAA model can have a mobility edge and an intermediate regime between localised and extended states. We next show in Sec.~\ref{sec:Bichromatic} that the presence of mobility edges and critical states is not unique to our choice of parameters in the 2D GAA model, with an example of a bichromatic potential.  We then consider a few example cases to study the presence of mobility edges in the spectrum and show that there are anomalous mobility edges in the 2D GAA model between both localised and critical states and extended and critical states. Finally, in Sec.~\ref{sec:Diffusion}, we consider the impact of the presence of critical states and mobility edges on the dynamical properties of initially localised states.

\section{The 2D Generalised Aubry-Andr\'e Model and properties}

\subsection{Model}

We consider the single-particle physics of a family of 2D on-site quasiperiodic tight-binding models given by the Hamiltonian
\begin{equation}
\begin{aligned}
H =  - J \sum_{\langle i,j \rangle} \left( \hat{b}^\dagger_i \hat{b}_j + \mathrm{h.c.} \right) - \lambda \sum_i V_{\beta}(i) \hat{n}_i,
\end{aligned}\label{eq:Ham}
\end{equation}
with tunnelling coefficient $J$, on-site modulation strength $\lambda$, on-site potential deformation parameter $\beta$, $\hat{b}^\dagger_i$ ($\hat{b}_j$) being the creation (annihilation) operator of a particle at the $i$th site, $\hat{n}_i$ the number operator, and $\langle i,j \rangle$ denoting nearest-neighbours.  We define the state of the system $\ket{\Psi}$ as
\begin{equation}
\ket{\Psi} = \sum_{i} \psi(x_i,y_i) \ket{i},
\end{equation}
where $\ket{i}$ labels the state with a single particle occupying the site labelled by $(x_i,y_i)$ with $ \psi(x_i,y_i)$ the coefficients that fully define the single-particle state, and which we will refer to as the wave function or state. We will consider the geometry of the lattice to be a standard 2D square lattice, though the general results do not rely on this choice of geometry.

We will consider the family of quasiperiodic models defined by an extension of the 1D GAA model \cite{Ganeshan2015Nearest} to 2D giving an on-site potential of
\begin{equation}\label{eq:GAA}
V_{\beta}(i) = \frac{V_{\mathrm{AA}}(x_i,y_i)}{1-\beta V_{\mathrm{AA}}(x_i,y_i)},
\end{equation}
with the 2D AA potential being
\begin{equation}
V_{\mathrm{AA}}(x_i,y_i) = \cos\left( 2 \pi (x_i+y_i) / \tau_1 \right) + \cos\left( 2 \pi (x_i-y_i) / \tau_2 \right)
\end{equation}
Moving forward, we will drop the $i$ label for each site and simply label sites via their $(x,y)$ coordinates in the lattice. By varying $\beta \in (-0.5,0.5)$ in Eq.~\eqref{eq:GAA}, a family of quasiperiodic potentials is explored as long as the periods $\tau_1$ and $\tau_2$ are irrational.  We will consider an $L \times L$ lattice with open boundary conditions and by default take $\tau_1=\tau_2=\sqrt{2}$ with other irrational $\tau$ expected to explore similar physics.  We will consider lattices of length $L=60$ sites unless otherwise stated. The 2D AA model is retained at $\beta=0$, including the self-dual point governing the localisation transition.  The 2D GAA potential contains divergences for $\beta= \pm 0.5$, which we will not consider.  Note, it is also possible to couple different one-dimensional chains of the AA potential in order to build a 2D generalisation of the AA model \cite{Rossignolo2019Localization}, which will build similar structures to here. 

\begin{figure}[t]
	\centering
	\includegraphics[width=0.98\linewidth]{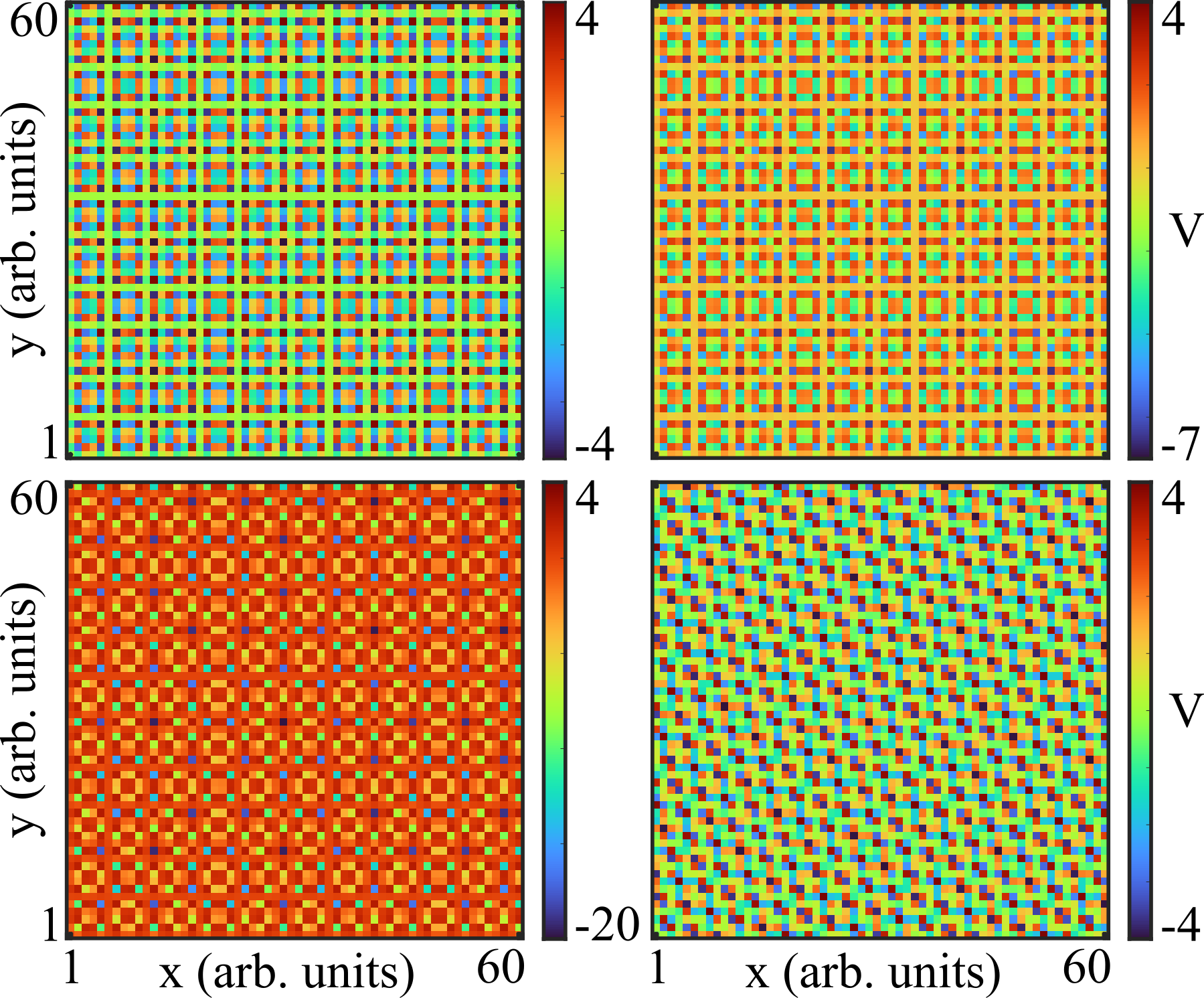}
	\caption{The potential of the 2D GAA model on a square lattice for $\lambda/J = 2$. The case of equal irrational periods of $\tau_1=\tau_2=\sqrt{2}$ with a deformation of (a) $\beta=0$, (b) $\beta=0.2$, and (c) $\beta-0.4$. The case of unequal irrational periods of $\tau_1=\sqrt{2}$ and $\tau_2=(1+\sqrt{5})/2$ is shown in (d) with a deformation of $\beta=0$.}
	\label{fig:pot}
\end{figure}

Examples of the 2D GAA model for $\tau_1 = \tau_2$ and an example for $\tau_1 \neq \tau_2$ are shown in Fig.~\ref{fig:pot}. In the case of $\tau_1 = \tau_2$ in Figs.~\ref{fig:pot}(a) to (c), weak modulation lines can be clearly observed, where the potential varies little across a line through the lattice. We also observe that in the case of $\tau_1 \neq \tau_2$ in Fig.~\ref{fig:pot}(d) there are still weak modulation regions that percolate through the lattice. However, when the symmetry of $\tau_1 = \tau_2$ is removed, the weak modulation does not appear along lines which match the geometry of the underlying square lattice. We will show that critical states rely on the presence of these weak modulation regions but do not fundamentally require  these regions to appear in the form of lines along the square lattice.

We briefly note that the 2D GAA model can be implemented in current cold atom set-ups in a similar way to the 1D GAA model \cite{Li2015Many},  i.e. by considering the limit of $\beta \ll 1$ as expanding $V_{\beta}(x,y)$ around $\beta=0$ gives
\begin{equation}
V_{\beta}(x,y) \approx V_{\mathrm{AA}}(x,y) + \beta \left[V_{\mathrm{AA}}(x,y)\right]^2 + \mathcal{O}(\beta^2).
\end{equation}
The additional term proportional to $\beta$ can be realised through the introduction of an additional lattice field along the direction of the square lattice and the AA modulation. There have also been proposals to realise the 1D GAA model via photonic lattices \cite{Liu2020Generalized}, and the 2D GAA model could be considered in a similar setting.

\subsection{Participation Ratios}\label{sec:PR}

The inverse participation ratio is a measure of the localisation of a quantum state within its Hilbert space. We will consider the inverse participation ratio
\begin{equation}
\mathrm{IPR}_n = \sum_{x,y} |\psi(x,y)|^4 ,
\end{equation}
and the normalised participation ratio
\begin{equation}
\mathrm{NPR}_n = \left(L^2 \sum_{x,y} |\psi(x,y)|^4\right)^{-1} ,
\end{equation}
of each state. We will also average over all states to gain an insight into the overall behaviour of the system and we will denote this by $\langle \mathrm{IPR} \rangle$ and $\langle \mathrm{NPR} \rangle$, with $\langle \cdot \rangle$ denoting the average over all states. We can utilise the participation ratios to determine how localised the states are within the Hilbert space, with localised states defined by $\langle \mathrm{IPR} \rangle \sim \mathcal{O}(1)$ and $\langle \mathrm{NPR} \rangle \sim 0$ and extended states by $\langle \mathrm{IPR} \rangle \sim 0$ and $\langle \mathrm{NPR} \rangle \sim  \mathcal{O}(1)$. A third intermediate regime where the spectrum is partially localised or extended can be defined when both $\langle \mathrm{IPR} \rangle \sim  \mathcal{O}(1)$ and $\langle \mathrm{NPR} \rangle \sim  \mathcal{O}(1)$. Note, that this intermediate regime does not immediately imply the presence of critical states, and will occur when you have a mobility edge between localised and extended states in the spectrum.

\subsection{Fractal Dimensions}

Multifractal analysis of the wave function can distinguish between critical, localised, and extended states. This has been useful in characterising the states at the metal-insulator transition of Anderson localisation  \cite{Castellani1986,Holzer1991,Schreiber1991,Vasquez2008,Evers2008}, one-dimensional quasiperiodic models \cite{saul1988wavefunctions,Hiramoto1989,Mace2016,Wang2020}, and two-dimensional quasiperiodic tilings \cite{rieth1998numerical,Yuan2000}. 

To calculate a system's fractal dimensions, we first split the probability density into $N_l$ boxes of linear size $l$ with the system being of linear size $L$ and $d$-dimensional. The probability of finding the particle in the $k$th box is given by
\begin{equation}
\mu_k(l) = \sum_{i \in l^d} |\psi(x_i,y_i)|^2,
\end{equation}
where $i$ runs over all sites within the box. The $q$th moment of the probability measure is
\begin{equation}
P_q(l) = \sum_{k=1}^{N_l} \mu_k^q(l).
\end{equation}
Within a certain range of the box scaling $\kappa = l/L$ the moments will show a power law scaling with an exponent $\eta(q)$, that is
\begin{equation}
P_q(\kappa) \propto (\kappa)^{\eta(q)}.
\label{eq:Scaling}
\end{equation}
This exponent (often called the mass exponent) is defined as 
\begin{equation}
  \eta(q) =
  \begin{cases}
    d(q-1) & \text{extended} \\
    0 & \text{localised} \\
    D_q(q-1) & \text{critical},
  \end{cases}
  \label{eq:ScalingCases}
\end{equation}
where $d$ is the physical space dimension. For a critical state, the exponent is dependent upon the fractal dimension $D_q$. 

We will utilise the fractal dimension to characterise the critical states of the 2D GAA model by focusing on measuring the $q=2$ fractal dimension, as this is equivalent to the box-counting dimension often used for fractal structures and measures the spread of the wave function over the supporting Hilbert space. In this case, the Hilbert space is the physical lattice, meaning that $D_2$ is a measure of locality in space, with $D_2 = 0$ for localised states, $D_2 = d$ for extended states and $0 < D_2 < d$ for critical states.




\section{Localisation, Mobility Edges, and Critical States}\label{sec:Properties}

\subsection{The Intermediate Regime}

We will now consider the localisation properties of the 2D GAA model. It has been shown that the 1D GAA and other 1D quasiperiodic models have an intermediate regime where localised and extended states coexist \cite{Li2020Mobility}.  In 1D, this intermediate regime is related to the presence of at least one mobility edge in the spectrum between localised and extended states and is identified by both the mean inverse and normalised participation ratios being non-zero.

It is known that the 2D Aubry-Andr\'e model (the case of $\beta=0$) is self-dual \cite{Szabo2020Mixed} and that this occurs at $\lambda = 2J$. The model then has extended states for $\lambda<2J$, localised states for $\lambda>2J$, and critical state at the transition point $\lambda=2J$. However, it was observed recently that this model also exhibits states that are partially extended, i.e. critical, over a large range of $\lambda$ away from the transition \cite{Szabo2020Mixed}.  However, the presence of partially extended or critical states does not guarantee that there is a mobility edge between localised and extended states or that there is a mix of localised and extended states at any given $\lambda$. 

\begin{figure}[t]
	\centering
	\includegraphics[width=0.98\linewidth]{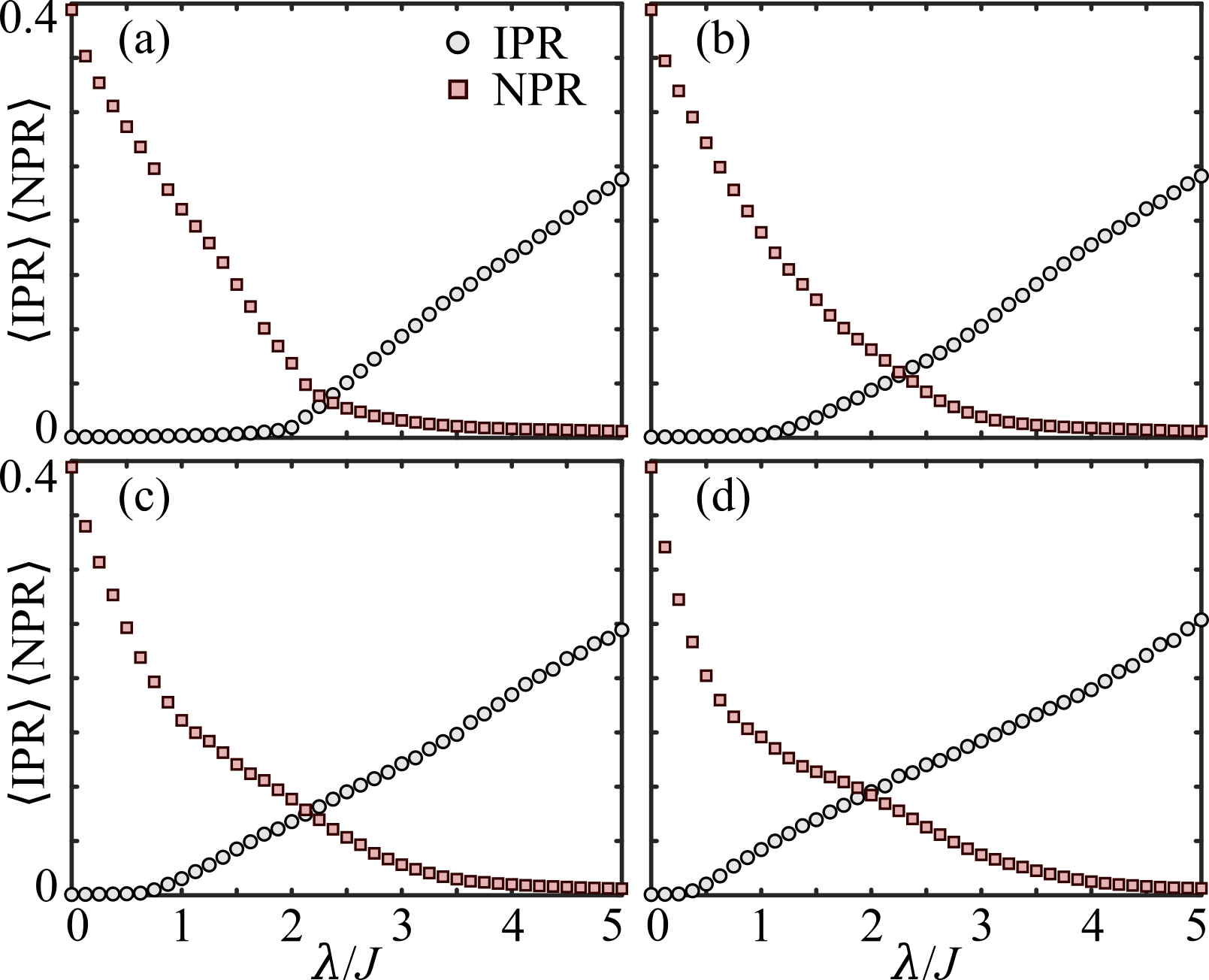}
	\caption{The mean over all states of the inverse and normalised participation ratios for the 2D GAA model. Shown are the cases of (a) $\beta=0$ (the 2D Aubry-Andr\'e model), (b) $\beta=0.2$, (c) $\beta=0.3$, and (d) $\beta=0.4$. The intermediate regime, where both $\langle \mathrm{IPR} \rangle$ and $\langle \mathrm{NPR} \rangle$ are larger than zero, is shown to grow with increasing $\beta$. }
	\label{fig:IPRNPR}
\end{figure}

We observe in Fig.~\ref{fig:IPRNPR}(a) that the 2D AA model has only a small region after the self-dual point where there is both a finite $\langle \mathrm{IPR} \rangle$ and $\langle \mathrm{NPR} \rangle$ and there is unlikely to be a significant mobility edge across a range of $\lambda$ between localised and extended states. As we increase the deformation parameter $\beta$ in Figs.~\ref{fig:IPRNPR}(b-d) we observe a broadening of the intermediate regime. This eventually reaches a point where the intermediate regime extends across a large range of modulation strengths for $\beta\geq 0.3$, as seen in Figs.~\ref{fig:IPRNPR}(c) and (d). With such a large region of mixed localised and extended states, it is expected that the system will contain at least one mobility edge that is being tuned with the deformation of the potential and stable with respect to $\lambda$. The 1D GAA model exhibits a similar tuning of the intermediate regime with $\beta$.

\begin{figure}[t]
	\centering
	\includegraphics[width=0.6\linewidth]{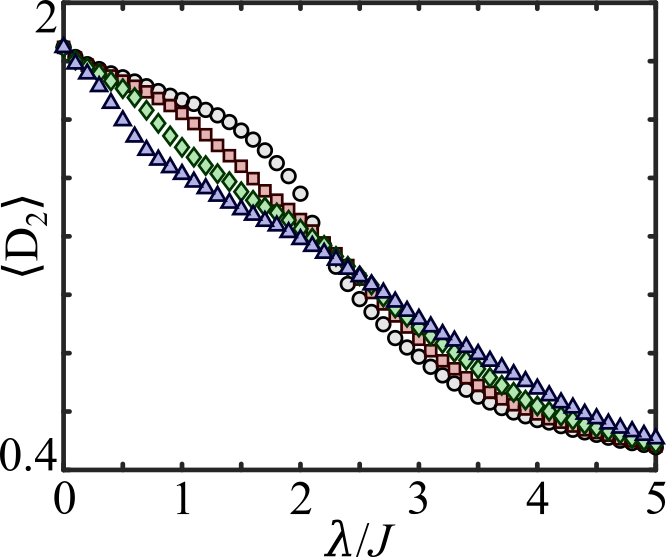}
	\caption{The mean scaling fractal dimension, $D_2$, of the 2D GAA model. Shown are the cases of $\beta=0$ by (black) circles, $\beta=0.2$ by (red) squares, $\beta=0.3$ by (green) diamonds, and $\beta=0.4$ by (blue) triangles. There is no clear signature of the intermediate regime changing in size as a function of $\lambda$ and the different quasiperiodic models show little difference in mean $D_2$.}
	\label{fig:meanD2}
\end{figure}

We also consider in Fig.~\ref{fig:meanD2} the mean scaling fractal dimension given by $\langle D_2 \rangle$ for a range of deformations $\beta$. If the majority of the states are either localised or extended, then we would expect to see a clear transition in the mean fractal dimension.  Then as we tune the modulation $\beta$ we would expect the transition to be `smoothed' due to the presence of a mix of extended and localised states with a corresponding mobility edge. We do observe in Fig.~\ref{fig:meanD2} the expected smoothing of the transition in $\langle D_2 \rangle$ due to the presence of a mobility edge as $\beta$ increases. However, we note that the transition in $\langle D_2 \rangle$ is already rather smooth and not sharp for the 2D AA model, but this model has been shown to have only a small intermediate regime in Fig.~\ref{fig:IPRNPR}(a).  The smoothing of the transition in $\langle D_2 \rangle$  with increasing deformation $\beta$ is also relatively minor compared to the large intermediate regions observed in Fig.~\ref{fig:IPRNPR}(b-d).  The fact that there are narrow and broad intermediate regimes but always a relatively smooth interpolation in $\langle D_2 \rangle$ may appear contradictory, but the reasons behind this apparent discrepancy will become clear as we consider the critical states that are present in Sec.~\ref{sec:critical}.

\begin{figure}[t]
	\centering
	\includegraphics[width=0.98\linewidth]{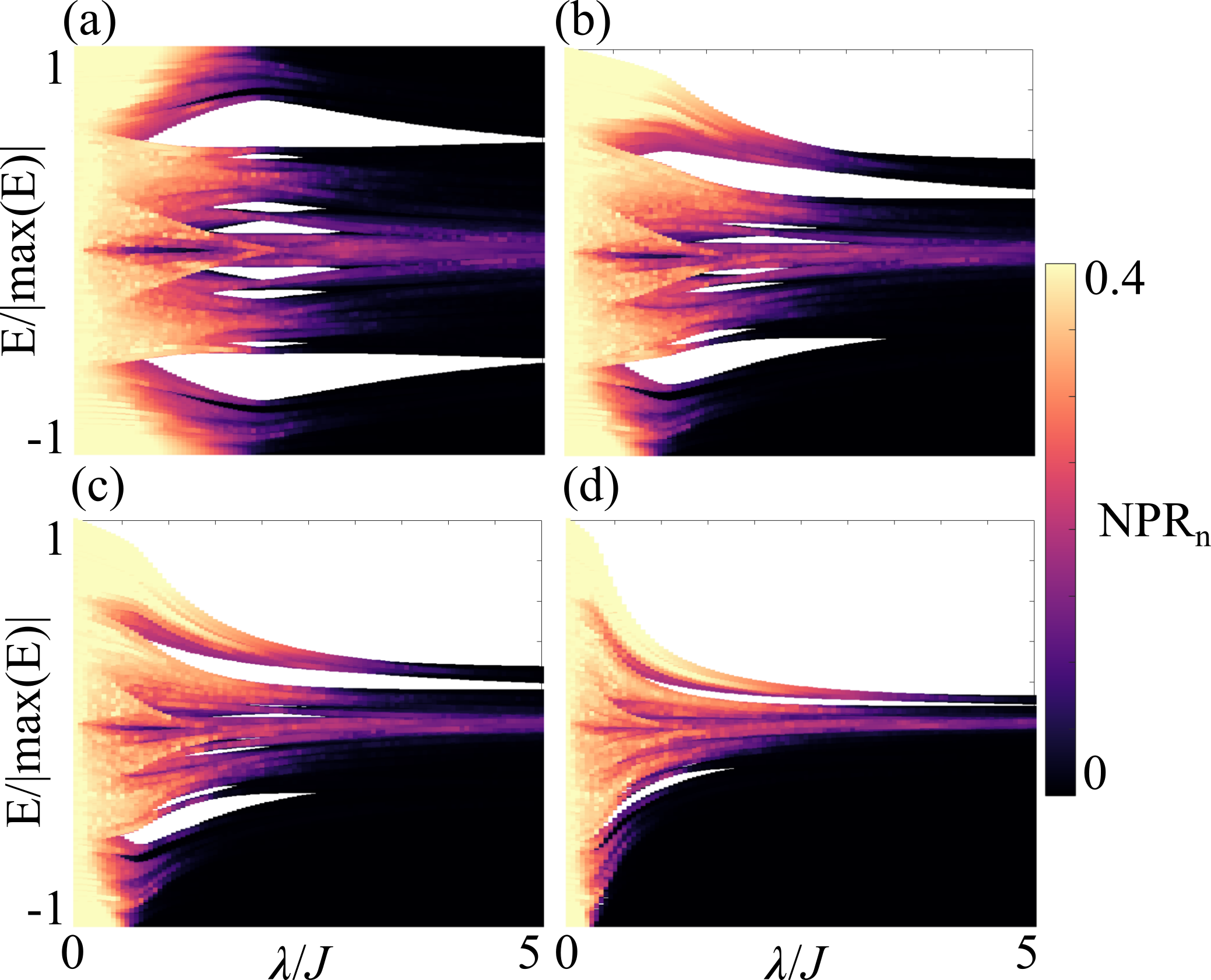}
	\caption{The NPR of all states for the 2D GAA model as a function of the potential strength $\lambda$. Shown are the cases for a deformation of the potential of (a) $\beta=0$, (b) $\beta=0.2$, (c) $\beta=0.3$, and (d) $\beta=0.4$. The deformation of the potential allows for the formation of a mobility edge between localised and extended states in the spectrum for some $\lambda$.}
	\label{fig:NPR}
\end{figure} 
 
\subsection{Mobility Edges}

We have shown that there exists a third intermediate mixed regime between localised and extended states for the 2D GAA model. The presence of such a regime has normally been expected to reflect a mixture of localised and extended states in the spectrum and, hence, the presence of at least one mobility edge.  We investigate the presence of mobility edges in the system by considering the individual $\mathrm{NPR}_n$ of each individual $n$th energy ordered state for a range of $\beta$ and across the localisation transition in $\lambda$, as is shown for four cases in Fig.~\ref{fig:NPR}. Note, similar results are obtained when considering the inverse participation ratio. First, we observe that the 2D AA model ($\beta = 0$) has no clear single-particle mobility edge between localised and extended states, with the majority of states $\mathrm{NPR} \rightarrow 0$ as we approach $\lambda/J=2$. As we turn on the deformation, we observe that a single-particle mobility edge is introduced, as shown in Figs.~\ref{fig:NPR}(b-c) with $\mathrm{NPR} \rightarrow 0$ faster for lower energy states.  The 2D GAA model therefore has a single-particle mobility edge between extended and localised states as a function of $\beta$ as is seen in the 1D GAA model. 

\begin{figure}[t]
	\centering
	\includegraphics[width=0.98\linewidth]{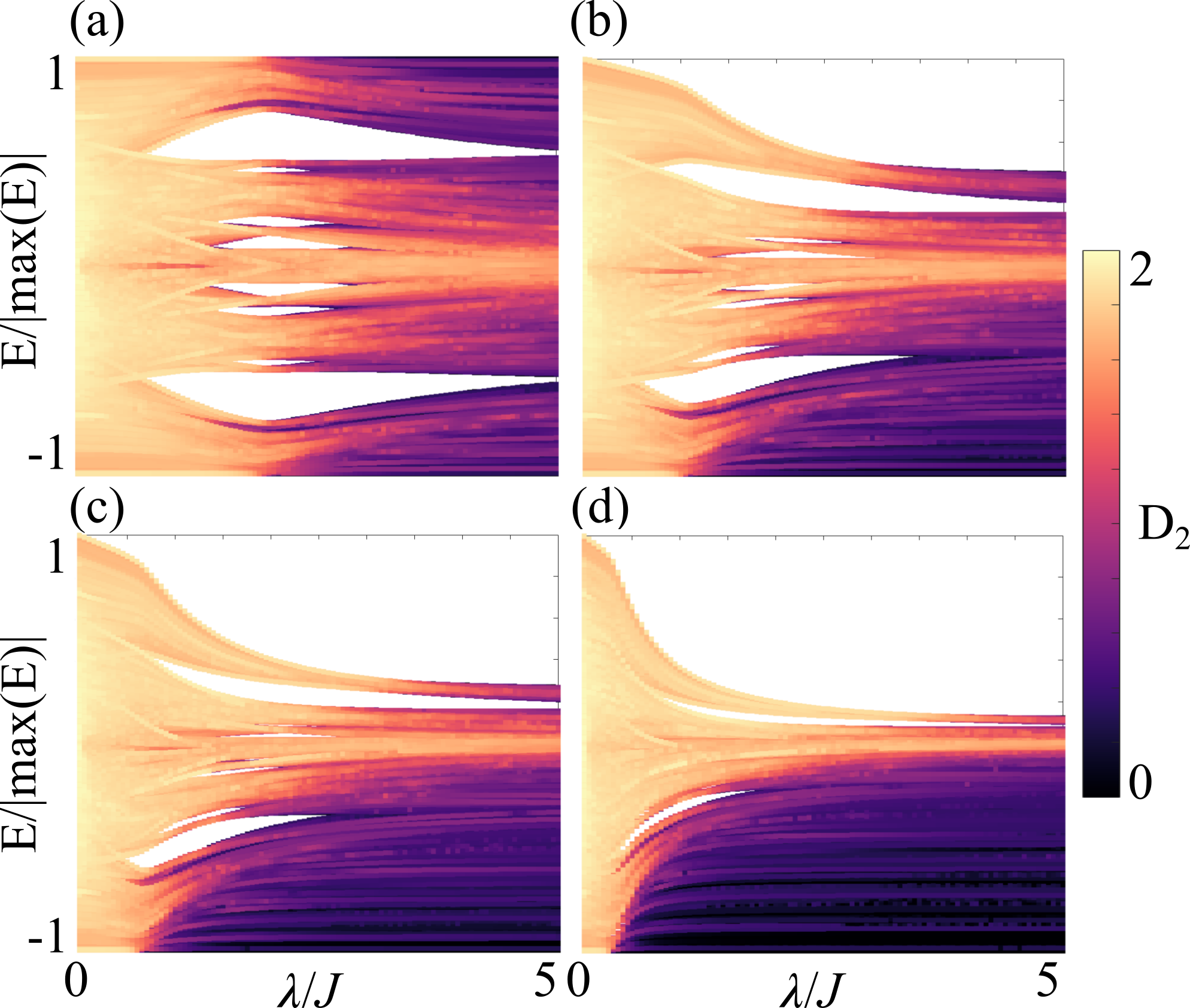}
	\caption{Same as Fig.~\ref{fig:NPR} but for the fractal scaling dimension $D_2$.}
	\label{fig:D2}
\end{figure}

\begin{figure}[t]
	\centering
	\includegraphics[width=0.98\linewidth]{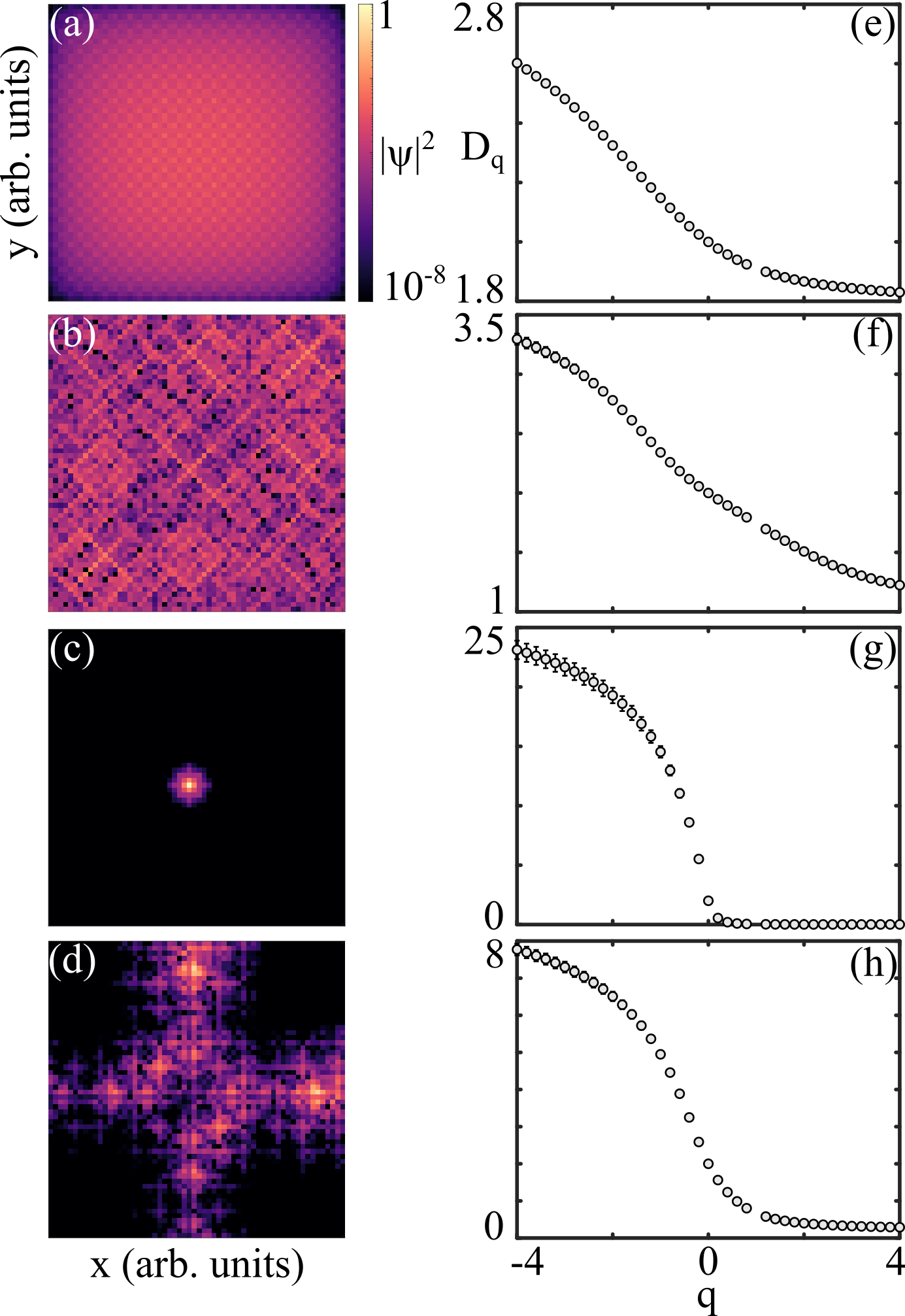}
	\caption{Examples of localised and extended states and critical states (a-b) on both sides of the localisation transition. We also plot the fractal dimensions (e-h) of the $q$th moment.  On the extended side of the transition we consider (a,e) an extended state and (b,f) a critical state.  On the localised side of the transition we consider (c,g) a localised state and (d,h) a critical state.}
	\label{fig:ExampleStates}
\end{figure}

\begin{figure}[t]
	\centering
	\includegraphics[width=0.98\linewidth]{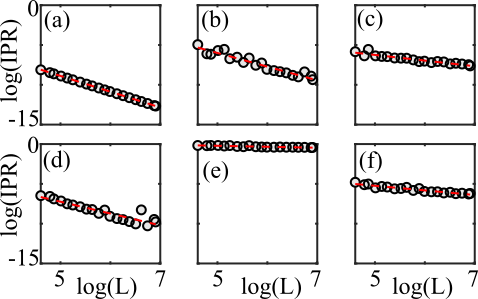}
	\caption{Scaling of the mean IPR over all states within different energy windows for (a-c) $\lambda=J$ and (d-f) $\lambda=3J$ for systems of linear size between $L=100$ ($\log 100 = 4.6$) and $1000$ ($\log 1000 = 6.9$). Linear fits to obtain $\gamma$ are shown by a dashed (red) line in each plot. (a) Energy window $ \bar{E} = E/|\mathrm{max}(E)| \in [-1,-0.9976]$ with $\gamma=-1.98\pm0.004$, (b) $\bar{E} \in [-0.0024,0]$ with $\gamma=-1.72\pm0.24$, (c) $\bar{E} \in [-0.0571,-0.0548]$ with $\gamma=-0.71\pm0.12$, (d) $\bar{E} \in [-0.0015,0]$ with $\gamma=-1.49\pm0.31$, (e) $\bar{E} \in [-1,-0.9985]$ with $\gamma=-0.12\pm0.03$, (f) $\bar{E} \in [-0.1542,-0.1527]$ with $\gamma=-0.58\pm0.09$.}
	\label{fig:IPRLarge}
\end{figure}

However, in all cases of $\beta$ in Fig.~\ref{fig:NPR} there appears to be some moderately extended states with a finite $\mathrm{NPR}_n$ for even large $\lambda$ centred around $E\sim 0$. These states are considerably localised in the 2D system, with  $\mathrm{NPR}_n$ being small compared to the values in the extended regime but also noticeably resistant to converging to  $\mathrm{NPR}_n = 0$ as would be expected for a localised state far into the localised regime. We propose that what is being observed in the normalised participation ratio is the presence of critical states in the system. However, the normalised participation ratio alone is not sufficient to confirm this, therefore we consider in Fig.~\ref{fig:D2} the scaling fractal dimension $D_2$ of each state again in energy ordering and across the extended to localised transition in $\lambda$ for the same $\beta$ as Fig.~\ref{fig:NPR}. With this we observe in all scenarios the presence of states with $D_2$ being consistent with the state being critical, i.e. neither $D_2\rightarrow d$ or $D_2 \rightarrow 0$. We note that finite size effects impact the ideal convergence of $D_2$ and we have checked for individual states that it converges with increasing system size to $d$ or $0$ for extended and localised states respectively. We note also, that for some $\lambda$ that would be expected to be in the localised regime, critical states dominate the central portion of the spectrum. Such regions could be expected to effectively remain in the extended or thermalising phase, as the presence of many critical states will support the delocalisation of a quantum state through the system, we will investigate this further in Sec.~\ref{sec:AME}. 

\subsection{Critical States}\label{sec:critical}

We will briefly consider some examples of critical states on both sides of the localisation transition. From the previous consideration of the scaling fractal dimension $D_2$, it can be seen that there are many states with intermediate $D_2$. Starting with the extended side of the localisation transition, we consider the case of $\lambda/J = 1$. In the considered $60 \times 60$ system, there are far too many states to consider all of them and we focus on two typical examples from each side of the transition. We first show a typical extended state in Fig.~\ref{fig:ExampleStates}(a) with its corresponding fractal dimensions in Fig.~\ref{fig:ExampleStates}(e). From inspection, the state is homogeneous and extended throughout the lattice. The fractal dimensions are also all $D_q\sim 2$, with deviations coming from the finite size of the system. A typical critical state for $\lambda<2J$ is shown in Fig.~\ref{fig:ExampleStates}(b) with its corresponding fractal dimensions in Fig.~\ref{fig:ExampleStates}(f). The fractal dimensions make the nature of this state clear, with it varying across the moments $q$, which is a signature of a critical state.

Moving to the localised side of the transition, we consider the case of $\lambda/J=4$. First, we show a typical localised state in Fig.~\ref{fig:ExampleStates}(c), this is heavily localised and its corresponding fractal dimensions tend to zero for positive moments, and effectively a numerical infinity ($\gg 2$) for negative moments as shown in Fig.~\ref{fig:ExampleStates}(g). A typical critical state on the localised side of the transition is shown in Fig.~\ref{fig:ExampleStates}(d) and it has a far more clear multifractal nature than that for the extended regime. There are regions of the state that are extended along some directions and localised along others. When looking at the fractal dimension for this critical state in Fig.~\ref{fig:ExampleStates}(h) it is large for negative $q$ and is non-zero but small for positive $q$, this reflects the localised but extended nature of this critical state.

We now want to ensure that the observed critical states are not a finite-size effect and will probe the scaling of the mean IPR of all states within specific small energy windows for increasing system size. The expected value of the IPR will scale with system size as $\mathrm{IPR} \sim L^\gamma$, with $\gamma=0$ for localised states, $\gamma=-d$ for extended states, and intermediate values being a signature of critical states \cite{Hiramoto1989}. We show examples of the scaling of the IPR for various energy windows for both the extended $\lambda=J$ and localised $\lambda=3J$ regimes in Fig.~\ref{fig:IPRLarge} by varying the system size from $10^2$ to $10^3$ sites. This includes clear examples of critical states for the extended regime in Fig.~\ref{fig:IPRLarge}(c), with $\gamma=-0.71\pm0.12$ in the energy window of $E/|\mathrm{max}(E)| \in [-0.0571,-0.0548]$, and the localised regime in Fig.~\ref{fig:IPRLarge}(f), with $\gamma=-0.58\pm0.09$ in the energy window of $E/|\mathrm{max}(E)| \in [-0.1542,-0.1527]$.

\subsection{Bichromatic Potential}\label{sec:Bichromatic}

\begin{figure}[t]
	\centering
	\includegraphics[width=0.98\linewidth]{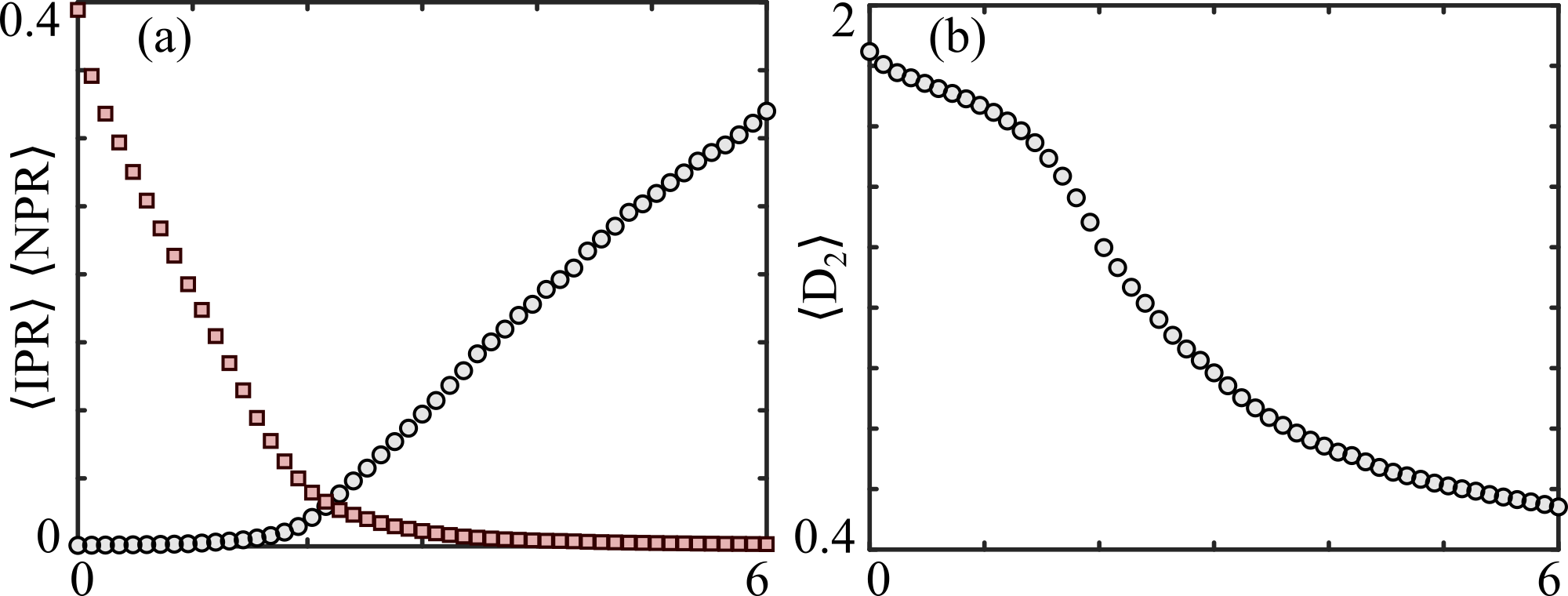}
	\caption{Mean properties of a bichromatic potential with $\tau_1=\sqrt{2}$ and $\tau_2=(1+\sqrt{5})/2$ in the 2D GAA model with deformation $\beta=0$. (a) The mean over all states of the inverse and normalised participation ratios. (b) The mean over all states of the scaling fractal dimension, $D_2$.}
	\label{fig:taus2}
\end{figure}

\begin{figure}[t]
	\centering
	\includegraphics[width=0.98\linewidth]{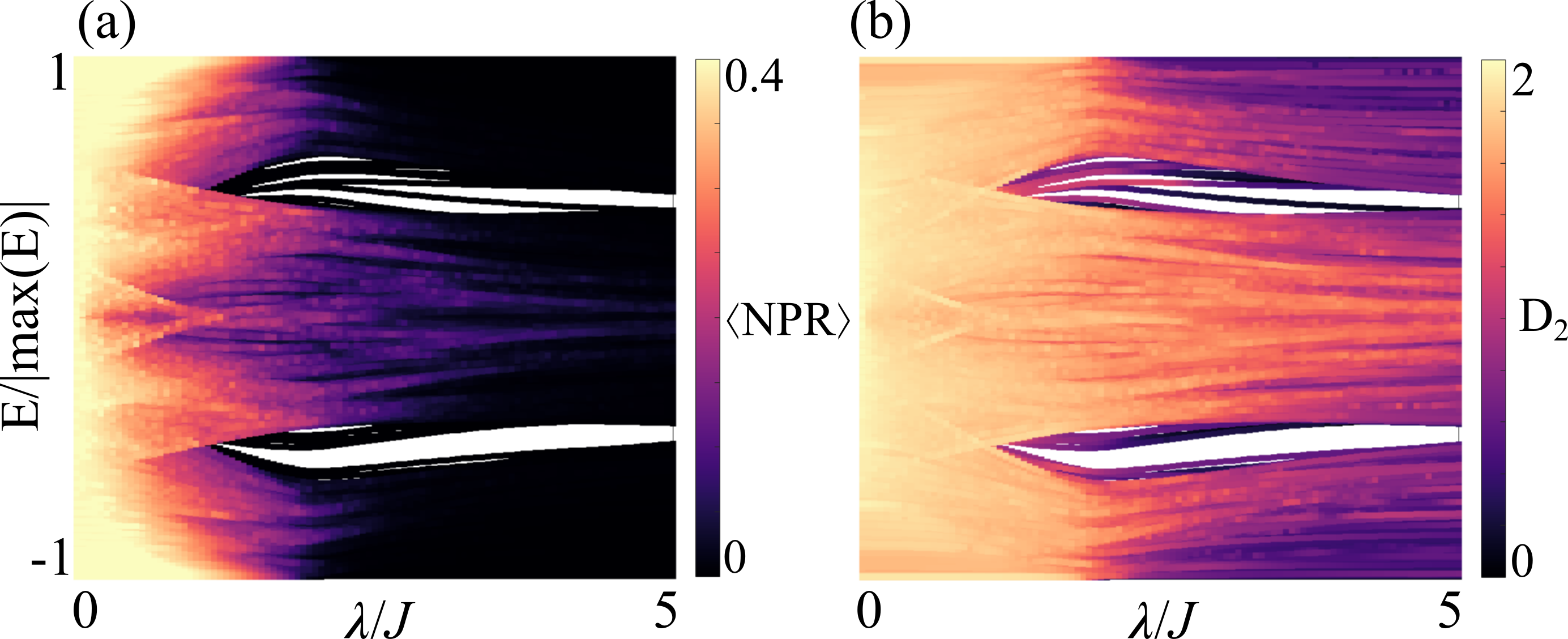}
	\caption{Properties of a bichromatic potential with $\tau_1=\sqrt{2}$ and $\tau_2=(1+\sqrt{5})/2$ in the 2D GAA model with deformation $\beta=0$. (a) The normalised participation ratios. (b) The scaling fractal dimension, $D_2$.}
	\label{fig:taus2all}
\end{figure}

\begin{figure*}[t]
	\centering
	\includegraphics[width=0.8\linewidth]{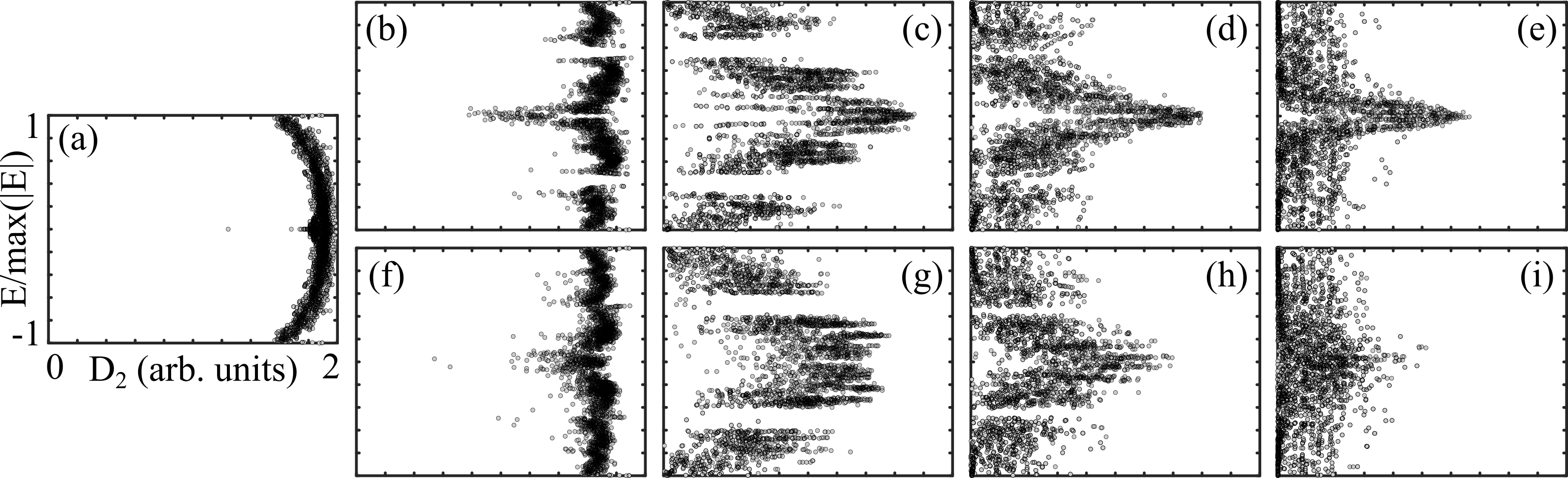}
	\caption{Examples of the fractal scaling dimension $D_2$ for each eigenstate of energy $E$ in the 2D GAA model. (a) Shows the trivial case of $\lambda/J=0$, this gives a reference for the effect of the finite size of the system on $D_2$. (b-e) Show the case of $\tau_1=\tau_2=\sqrt{2}$, and (f-i) the case of  $\tau_1=\sqrt{2}$ and $\tau_2 = (1+\sqrt{5})/2$. From left to right shows increasing modulation strength of (b,f) $\lambda/J=1$, (c,g) $\lambda/J=3$, (d,h) $\lambda/J=5$, and (e,i) $\lambda/J=10$. It is clear that in all cases of $\lambda\neq 0$ the spectrum includes critical states and that there are mobility edges between extended and critical states as well as localised and critical states.}
	\label{fig:D2SameTau}
\end{figure*}

The observation of critical states in the 2D GAA model with $\tau_1=\tau_2$ poses an important question, are the critical states present a direct result of the weak modulation lines matching the geometry of the tight-binding model, i.e. the weak modulation is along a direction which is connected by tunnelling. As shown in Fig.~\ref{fig:pot}, when $\tau_1\neq\tau_2$ the weak modulation lines do not follow the geometry of the underlying lattice. We now consider the case of $\tau_1 = \sqrt{2}$ and $\tau_2 = (1+\sqrt{5})/2$ to show that the exact form of the weak modulation lines does not change the presence of critical states.  As the 2D GAA potential is now defined by two periods (or frequencies) we will refer to the potential as being bichromatic. We will consider only the case of $\beta=0$, with non-zero $\beta$ resulting in similar effects to what has already been observed, e.g., the introduction of a mobility edge between extended and localised states and a larger intermediate regime as a result.

We first confirm that the bichromatic potential has an extended to localised transition, as shown by the $\langle \mathrm{IPR} \rangle$ and $\langle \mathrm{NPR} \rangle$ in Fig.~\ref{fig:taus2}(a). We also see the same behaviour as the chromatic case for the mean fractal scaling dimension $\langle \mathrm{D_2} \rangle$ in Fig.~\ref{fig:taus2}(b). As shown in Fig.~\ref{fig:taus2all}(a), there is an extended to localised transition of all states with no obvious mobility edge in the spectrum. Though, if $\beta\neq 0$, a single particle mobility edge between extended and localised states is still observed. Importantly we show that the bichromatic potential still supports critical states, as shown by the large region of $0 < \langle \mathrm{D_2} \rangle < 2$ in Fig.~\ref{fig:taus2all}(b).

\section{Anomalous Mobility Edges and Diffusion}\label{sec:AME}

\subsection{Mobility Edges}

We will now consider in more detail the nature of the mobility edges present and show that there are mobility edges in all cases of the 2D GAA model.  This is due to the presence of anomalous mobility edges between extended or localised and critical states in the spectrum, even deep into the extended or localised regime. To see this we plot the scaling fractal dimension as a function of the eigenenergies for the full spectrum for a number of different values of $\lambda$.  As a reference point for the influence of the finite size and underlying tight-binding model without a quasicrystalline potential, we plot the case of $\lambda=0$ in Fig.~\ref{fig:D2SameTau}(a). Here we observe that the majority of the states have $D_2 \approx 2$, with deviations being evident for the lowest and highest energy states where the finite size of the system is particularly impactful.  There is also a noticeable deviation from $D_2 \approx 2$ at the centre of the spectrum, this is due to the properties of states in the central flat-band which is typical in 2D noninteracting tight-binding models.

First, we consider the case of equal irrational periods in the 2D GAA potential, with $\tau_1=\tau_2=\sqrt{2}$ in Fig.~\ref{fig:D2SameTau}(b-e). For $\lambda=J$ as shown in Fig.~\ref{fig:D2SameTau}(b), the potential causes the states in the central portion of the spectrum and elsewhere to localise with $D_2 \approx 1$, i.e. their support is effectively that of a one-dimensional system.  Therefore, for $\lambda < 2J$ there are at least two mobility edges between critical and extended states. Past the localisation transition the edges of the spectrum show the presence of localised states with $D_2 \approx 0$ but the central portion of the spectrum remains critical. The complexity of defining mobility edges and when states are extended, critical, or localised becomes evident in Fig.~\ref{fig:D2SameTau}(c) and (d), as states of similar energy can be characterised by vastly different $D_2$. It is clear that anomalous mobility edges between localised and critical states will be present and there could even also be some between extended and critical states for modulations close to the critical point, as in Fig.~\ref{fig:D2SameTau}(c). When the modulation becomes strong, as is shown in Fig.~\ref{fig:D2SameTau}(e) for $\lambda=10J$, then the majority of states become localised. However, it is clear that a large portion of the states in the central region of the spectrum remain critical and there will still be an anomalous mobility edge between localised and critical states.  We can confirm that while this central region of critical states does get narrower in energy, it is present even for large modulations up to and including $\lambda=100J$, see Sec.~\ref{sec:Diffusion}.

\begin{figure}[t]
	\centering
	\includegraphics[width=0.95\linewidth]{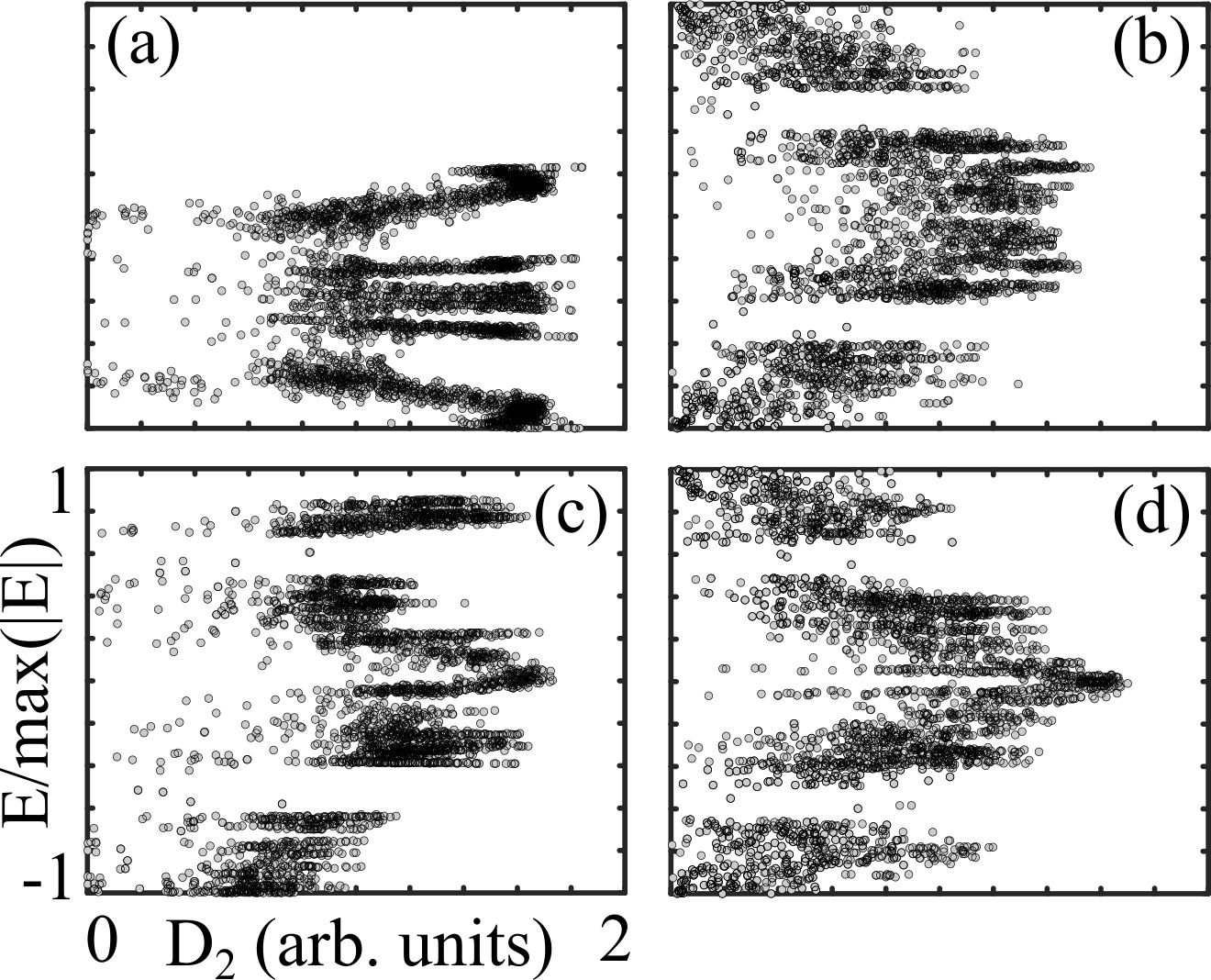}
	\caption{Examples of the fractal scaling dimension $D_2$ for each eigenstate of energy $E$ in the 2D GAA model for fixed $\tau_1=\sqrt{2}$ and $\lambda/J=3$ with varying $\tau_2$. Shown are examples of (a) $\tau_2=\sqrt{1}$, (b) $\tau_2=\sqrt{3}$, (c) $\tau_2=\sqrt{9}$, and (d) $\sqrt{10}$. Examples (a,c) are therefore crystalline in a single direction as the potential is a set of repeating 1D quasiperiodic chains, and this is reflected in the number of states with $D_2\approx 1$ shown in each case.}
	\label{fig:D2OtherTau}
\end{figure}

We now consider the case of unequal irrational periods in the 2D GAA potential, with $\tau_1=\sqrt{2}$ and $\tau_2=(1+\sqrt{5})/2$ in Fig.~\ref{fig:D2SameTau}(f-i). The results are very similar to that of the equal periods already considered. Especially for modulation strengths around the transition, as shown in Fig.~\ref{fig:D2SameTau}(g) and (h).  This shows again that the presence of critical states and anomalous mobility edges in the spectrum is not reliant on their being weak modulation lines formed that match the geometry of the tunnelling in the tight-binding model. However, the extreme cases of $D_2$ for both low and high modulation strengths are not present as shown in Fig.~\ref{fig:D2SameTau}(f) and (i). We also consider the case of fixed $\tau_1=\sqrt{2}$ with $\tau_2=\sqrt{3}$ and $\tau_2=\sqrt{10}$ in Fig.~\ref{fig:D2OtherTau}(b) and (d) respectively.  In both cases, we again observe a similar structure in the distribution in $D_2$ to the cases already discussed in Fig.~\ref{fig:D2SameTau}. 

We now consider two examples to illustrate the origins of the critical states by having one of the periods be rational. In these examples the model is quasiperiodic only in one direction (that of $(x+y)$ in our case) and periodic or crystalline in the other. We take the case of $\tau_1=\sqrt{2}$ with $\tau_2 = \sqrt{1}$ and $\tau_2=\sqrt{9}$ in Fig.~\ref{fig:D2OtherTau}(a) and (c) respectively.  We again observe a range of different scaling fractal dimensions for states throughout the spectrum.  The origin of the critical states for the superlattice setting is particularly clear in Fig.~\ref{fig:D2OtherTau}(c) with a large number of states having $D_2 \approx 1$, i.e. they are one-dimensional states and this is due to the system now being effectively a set of coupled 1D chains of tight-binding models, as was used to build the 2D AA model from stacking AA chains \cite{Rossignolo2019Localization}. Interestingly, we observe similar properties of the states in the bichromatic quasiperiodic case shown in Fig.~\ref{fig:D2SameTau}(g).

\subsection{Diffusion}\label{sec:Diffusion}

We have so far shown that critical states play a central role in the spectral properties of the 2D GAA model, and that this results in the presence of anomalous mobility edges between critical states and localised or extended states in the spectrum. We will now investigate how this impacts the physical properties of the 2D GAA model by considering the diffusion properties of initially localised states. The state will be initialised in a single site and as the model is quasiperiodic through the on-site potential, this means that starting at different initial sites will be equivalent to starting at different energies. By changing the initial site that is populated we can then sweep through the spectrum, allowing us to probe the impact of the mobility edges present on the diffusion of an initially localised state.  The initial state will be propagated under the time-independent Hamiltonian described in Eq.~\eqref{eq:Ham} for a fixed total duration of time in the first instance. We will specifically study the same $60\times 60$ system with open boundaries that we have focused on to this point and evolve for a time $t=100J^{-1}$.  In order to measure the diffusion of the particle through the system we will measure the mean square deviation
\begin{equation}
\sigma_x = \sqrt{\langle x^2 \rangle - \langle x \rangle^2},
\end{equation}
where $\langle x \rangle$ is the expectation in $x$ of the state. Note, we will consider the case of $\beta=0$ in detail and focus on the spread in a single dimension $\sigma_x$ with similar results observed for $\sigma_y$.

\begin{figure}[t]
	\centering
	\includegraphics[width=0.95\linewidth]{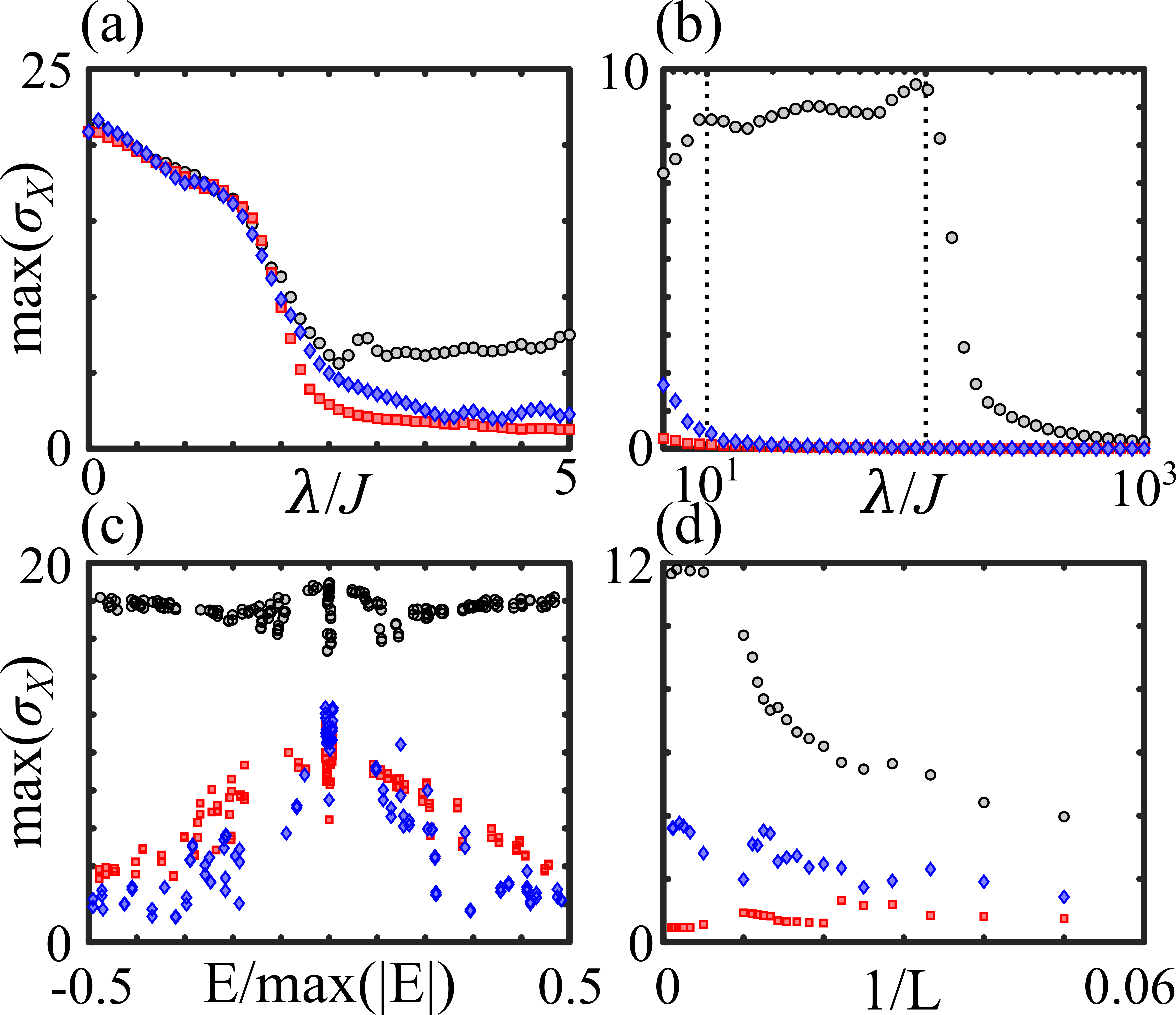}
	\caption{Anomalous diffusion in the 2D GAA model for $\beta=0$, and $\tau_1=\tau_2=\sqrt{2}$. (a,b) Show the maximum mean square displacement across potential strengths $\lambda/J$ after $t=100J^{-1}$ for states initially localised to a single site. Initial states of different energy are shown, with squares (red) showing a state with energy near the ground state, diamonds (blue) showing a state at approximately half of the ground state, and circles (black) showing a state with $\bar{E}\approx 0$. (c) Shows the distribution of maximum mean square displacement at various $\lambda$ after $t=100J^{-1}$, with circles (black) being $\lambda=J$, squares (red) being $\lambda=3J$, and diamonds (blue) being $\lambda=5J$. (d) Scaling of the maximum mean square displacement as a function of the side length $L$ with $\lambda=4J$ after $t=100J^{-1}$ and the points as described for (a,b).}
	\label{fig:sigma}
\end{figure}

In Fig.~\ref{fig:sigma}(a) we show the maximum mean square displacement across $\lambda$ for states that probe different parts of the spectrum. As expected, we observe that for all parts of the spectrum, the maximum mean square displacement starts off large for small $\lambda$, as the system is fully described by extended states. As $\lambda$ is tuned to higher values the extent of the state after propagation decreases, with a clear sharp transition at the AA delocalised to localised transition point of $\lambda=2J$. After this point, the different energy states begin to behave differently. The state initialised close to the ground state energy is converging to a small mean square deviation, a clear sign that the states around this energy are becoming more and more localised as has been reflected for the ground state in results throughout this work. However, states initialised close to the centre of the spectrum at $E\approx 0$,  at first follow this trend towards small mean square displacement but converge to a non-zero value for large $\lambda$ due to the presence of transport supporting critical states.  States between the ground state and centre of the spectrum largely follow the qualitative form of the ground state trend but with a clear extended transition in going to small $\sigma_x$ due to the presence of critical states. 

The middle and $E\approx 0$ state both converge to the small mean square displacement obtained for the ground state for large $\lambda$ as shown in Fig.~\ref{fig:sigma}(b) How this happens for the $E \approx 0$ states is particularly interesting, the critical states supporting the non-zero mean square displacement are stable well into large $\lambda$. However, this diffusion is eventually destabilised for dominating $\lambda$ of order $10^2J$, then at $10^3J$ the eigenstates are simply the Hilbert space, i.e. the occupation of each individual site. 

\begin{figure}[t]
	\centering
	\includegraphics[width=0.9\linewidth]{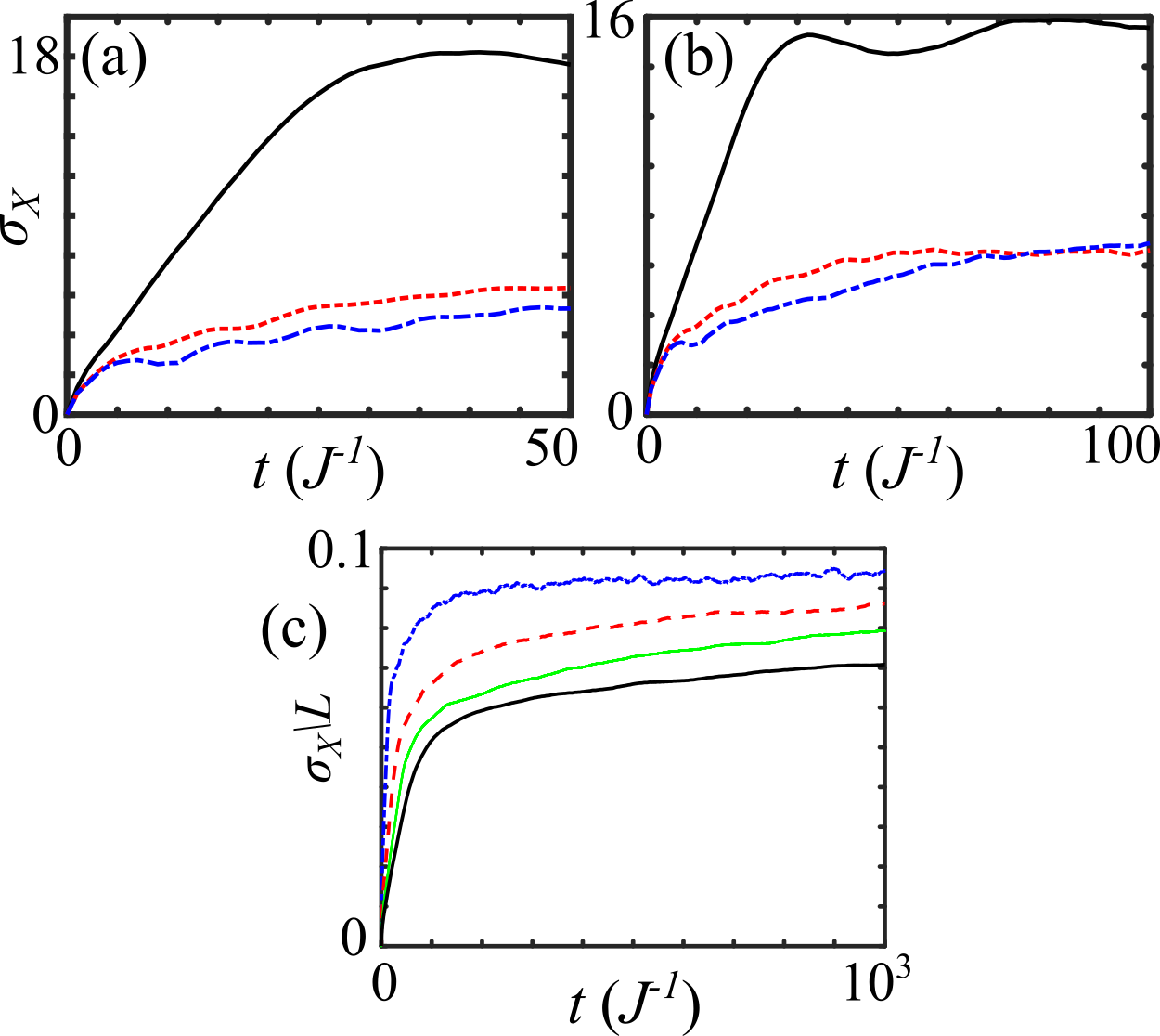}
	\caption{Expansion of localised states in the 2D GAA model with $\tau_1=\tau_2=\sqrt{2}$. Showing mean square displacement at each point in time as an initially localised state is propagated in time with the 2D GAA Hamiltonian. (a,b) Shown are the cases of $\lambda=J$ by a solid (black) line, $\lambda=3J$ by a dash-dot (blue) line, and $\lambda=4J$ by a dashed (red) line for an initial localised state with $\bar{E}\approx0$ for (a) $\beta=0$ and (b) $\beta=0.3$. (c) The case of $\lambda=4J$ for long time dynamics for large systems of size $L=250$ shown by a dash-dot (blue) line, $L=500$ by a dashed (red) line, $L=750$ by a dotted (green) line, and $L=1000$ by a solid (black) line.}
	\label{fig:time}
\end{figure}

The exhibited diffusion properties for the states are not unique to the three examples shown in Fig.~\ref{fig:sigma}(a) and (b) with Fig.~\ref{fig:sigma}(c) showing the spread of the maximum mean square displacement across the range of energies in the central portion of the spectrum for various $\lambda$.  We also confirm that this is not a finite-size effect by considering the scaling of the maximum mean square displacement up to $L=1000$ in Fig.~\ref{fig:sigma}(d), with a clear convergence for large systems giving a propagation limit in this fixed total time of propagation. We show an example of the long-time dynamics in Fig.~\ref{fig:time}(c), showing that for 2D lattices of up to $10^6$ sites, critical states can host diffusion of an initially localised state. Note, $\sigma_x/L$ is decreasing with increasing system size but even after $t=10^3J^{-1}$, the dynamical state is still slowly diffusing and the maximum $\sigma_x$ is not yet necessarily reached. Also, the number of possible starting sites for an initially localised particle to have similar energy and hence to reach this maximum $\sigma_x$ is also increasing.

Finally, we show the mean square displacement as time evolves for an initial state at $E\approx 0$ and localised to a single site in Fig.~\ref{fig:time} for both the case of $\beta = 0$ as discussed so far and for $\beta = 0.3$. For zero or non-zero $\beta$ the displacement for states at small lambda, $\lambda=J$ is shown, there is a clear linear transport regime, reflecting the ballistic expansion of the initially localised state into the space, followed by a saturation as the state has spread through the majority of the lattice. However, if we consider $\lambda$ across the standard AA localisation transition point, $\lambda=3J$ and $4J$ are shown, then there is a clear departure from the ballistic transport in favour of a more diffusive regime where the particle spreads slowly through the lattice. This reflects the results of Fig.~\ref{fig:sigma}, and shows that the non-zero mean square displacements observed across the central portion of the spectra and for a broad range of $\lambda$ are due to anomalous diffusion and the presence of critical states.

\section{Conclusions}

We have investigated the single-particle properties of two-dimensional models with quasicrystalline on-site terms through the 2D GAA model. It has been shown that similar physics to the 1D GAA model can be observed, with there being an intermediate regime and the deformation property of the potential introducing a mobility edge between extended and localised states. We have also shown that critical states are a general property of single-particle models in diagonal quasicrystalline models. This includes the support of critical states well within both the localised and extended regimes of the lattice and through a range of deformations $\beta$ and different combinations of the irrational periods $\tau$. These models also host anomalous mobility edges between both localised and critical states and extended and critical states. This especially impacts the transport properties of the microscopic model, with it possible to observe expansion through the presence of critical states.

There are a number of open questions that are raised from this work, with the key one being how the critical states impact the behaviour of many-body  quasicrystalline models. In particular, how does the presence of critical states impact the formation of states like the Bose glass, contribute to glassy dynamics, and  change the prospects of many-body localisation. In the latter case, it can be speculated that the critical states would thermalise any localised state in the long time limit. Initial results have shown promise of a stable many-body localised phase \cite{strkalj2022} but these are limited by relatively small system sizes for quasicrystalline systems with $\sim 100$ total sites.

With the current difficulty in pursuing full spectrum results or long-time dynamics in many-body numerical calculations, a route forward could be to study this system in a controlled experimental setting, e.g. with ultracold atoms in optical lattices \cite{lewenstein2012ultracold,Gross2017,schafer2020tools}. The 2D quasicrystalline optical potential realised in Ref.~\cite{Viebahn2019Matter} will contain critical states, and this has been investigated via numerics at the single-particle level \cite{Gottlob2023,zhu2023localization}. The geometry of the 2D GAA model considered in this work is a crystalline square lattice which can be realised through the generation of an optical lattice. The on-site modulation can then be realised by either a second rotated and weaker optical lattice, or through the manipulation of the individual lattice sites through digital mirror devices. The interaction of the atoms can then be controlled through Feshbach resonances and tuning the depth of the confining optical lattice, allowing for the realisation of the Bose-Hubbard model equivalent to the model considered here. The interactions can be tuned to zero and the physics discussed here could be observed. One can then envisage looking at increasing the interactions to attempt to observe the impact of the critical states on both the ground state and transport properties of the many-body system. While currently ambitious, this would be a similar experiment to that recently conducted to observe delocalisation mechanisms for 1D disordered systems with a quantum gas microscope \cite{leonard2023probing}.

\acknowledgements{The author thanks D. Johnstone, L. Sanchez-Palencia, E. Gottlob, and A. J. Daley for helpful discussions.  The author acknowledges support from DesOEQ EP/P009565/1.}



%

\end{document}